\documentclass[a4paper,twocolumn,11pt,accepted=2017-05-09]{quantumarticle}
\pdfoutput=1
\usepackage[utf8]{inputenc}
\usepackage[english]{babel}
\usepackage[T1]{fontenc}
\usepackage{amsmath}
\usepackage{hyperref}
\usepackage[utf8]{inputenc}
\usepackage[english]{babel}
\usepackage[T1]{fontenc}
\usepackage{csquotes}
\usepackage{hyperref}
\usepackage{tikz}
\usepackage{lipsum}

\usepackage{algorithm2e}
\usepackage[braket, qm]{qcircuit}
\usepackage{graphicx}
\usepackage{braket}
\usepackage{tikz}
\usetikzlibrary{shapes.geometric, arrows, matrix}
\usetikzlibrary{shapes, arrows}
\usepackage{pgfplots}
\usepgfplotslibrary{groupplots}

\usepackage{listings}
\usepackage{subcaption} 
\usepackage{algorithmicx}
\usepackage{float}
\usepackage{algpseudocode}
\usepackage{ifthen}
\usepackage{amssymb}
\usepackage{tabularx}
\usepackage{rotating}
\usepackage{multirow}
\usepackage{changepage} 
\pgfplotsset{width=8cm,compat=1.9}

\begin{document}

\title{A Generalized Space-Efficient Algorithm for Quantum Bit String Comparators}

\author{Khuram Shahzad}
\affiliation{Dept. of Computer Science, National University of Computer and Emerging Sciences, Peshawar, Pakistan}
\orcid{0000-0003-4059-1207}

\author{ Omar Usman Khan }
\affiliation{Dept. of Computer Science, National University of Computer and Emerging Sciences, Peshawar, Pakistan}
\orcid{0000-0003-2605-0623}


\maketitle

\begin{abstract}
Quantum Bit String Comparators (QBSC) operate on two sequences of n-qubits, enabling the determination of their relationships, such as equality, greater than, or less than. This is analogous to the way conditional statements are used in programming languages. Consequently, QBSCs play a crucial role in various algorithms that can be executed or adapted for quantum computers. The development of efficient and generalized comparators for any $n$-qubit length has long posed a challenge, as they have a high-cost footprint and lead to quantum delays. 

Comparators that are efficient are associated with inputs of fixed length. As a result, comparators without a generalized circuit cannot be employed at a higher level, though they are well-suited for problems with limited size requirements. In this paper, we introduce a generalized design for the comparison of two $n$-qubit logic states using just two ancillary bits. The design is examined on the basis of qubit requirements, ancillary bit usage, quantum cost, quantum delay, gate operations, and circuit complexity, and is tested comprehensively on various input lengths. The work allows for sufficient flexibility in the design of quantum algorithms, which can accelerate quantum algorithm development. 
\end{abstract}

\section{Introduction}
Quantum computing and its related algorithms have witnessed remarkable advances in recent years, owing to the principles of quantum mechanics and enhanced computing power. They offer the potential to efficiently solve many mathematical problems that are intractable for classical computers. For instance, Shor’s algorithm\cite{zhuang2022quantum} can achieve polynomial-time solutions for hard problems such as integer factorization or discrete logarithms. Similarly, Grover’s algorithm\cite{zhuang2022quantum} can provide a quadratic speedup for an unstructured search problem.

The quantum bit string comparator (QBSC) is a crucial component in quantum algorithms as it incorporates conditional statements, expanding the range of applications for quantum algorithms. It allows quantum programmers to utilize successful techniques from the classical computation that rely on comparisons \cite{oliveira2007quantum}. In a classical sense, a conditional statement leads to two mutually exclusive states. However, in the quantum domain, such conditional statements may lead to the merging of both branches due to superposition. Designing a comparator for quantum circuits is a significant challenge for researchers in the field. Existing quantum comparators are not scalable in terms of input size, as the circuit size depends on the number of inputs. Therefore, these quantum comparators, which lack a generalized circuit, are not suitable for higher-level applications, although they can handle problems with small size requirements. However, many applications, such as integer factorization, optimization, option pricing \cite{stamatopoulos2020option}, and risk analysis, often require one of the inputs to be classical.\cite{yuan2023improved} This necessitates a generalization of the circuit, especially when further computations are involved. 

Numerous approaches have been put forth to effectively devise quantum comparators. These include the serial-based approach \cite{wang2012design,al2009closed, oliveira2007quantum, xia2018efficient}, the tree-based approach \cite{thapliyal2010design, vudadha2012design} and the Quantum Fourier transform (QFT) \cite{yuan2023improved}. In the serial-based quantum comparator, comparisons of quantum bits are executed sequentially from the least significant bit to the most significant bit. Conversely, a tree-based quantum comparator can evaluate quantum bits for comparison in parallel. Although it holds an advantage over the serial-based approach in terms of time delay, it still falls short in terms of quantum cost. However, both approaches require one or two bits for comparison, along with two or more ancillary bits. 
This current study presents a generalized quantum comparator having a minimal quantum cost and optimized quantum delay that can compare two classical numbers of any size in binary form using a quantum circuit. The data on qubits remains unchanged and can also be used for other operations. Our quantum comparator has a linear scaling of Qubit resources with respect to the size of the input numbers. For example, it takes two ancillary qubits to compare two n-bit numbers. This is an improvement over some existing quantum comparators that require more ancillary qubits or have a higher quantum cost and higher quantum delays \cite{wang2012design, al2009closed, thapliyal2010design, vudadha2012design, oliveira2007quantum, xia2018efficient}.
The research analyzes the comparator qubit requirements, gate operations, and circuit complexity for various input sizes. Our proposed comparator's performance is comprehensively analyzed with respect to quantum cost, quantum delay, and ancillary bits. Since quantum comparators are fundamental in many quantum algorithms and applications; e.g. integer factorization, optimization, option pricing, and risk analysis, the work holds good promise for the design and development of quantum algorithms \cite{yuan2023improved}. 
\subsection{Quantum Gates for GQBSC}
Quantum gates are analogous to the application of various transformations on input states represented through qubits. In this section, we describe some fundamental gates that are used in GQBSCs. Each of the gates here performs some specific unitary operation, and as such, they are also represented using unitary matrices \cite{li2014multidimensional, li2013quantum}.

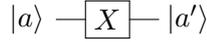
\begin{figure}
	\centering
	\scalebox{1}{
	{\Qcircuit @C=1em @R=1em {
			\ket{a}\qquad & \gate{X} & \qquad\ket{a'} \qw \\
		}
	}}
	\caption{NOT Gate\label{fig:notgate} representation for $\ket{a'} = X \ket{a}$}
\end{figure}
    
    
    
    
    
    

A single-qubit unitary case is that of the NOT gate, which is used for bit-flip operations, and is given in Fig.~\ref{fig:notgate} along with its mathematical representation. The gate represents the Pauli X operator and exhibits the properties $X^2 = I$, where I is the Pauli Identity operator. The bit-flip here is geometrically interpreted a half turn about the $x$-axis in a Bloch sphere.

\begin{figure}
	\centering
		\scalebox{1}{
		\Qcircuit @C=1em @R=1em {
		\ket{a}\qquad & \ctrl{1} & \qquad\ket{a}\qw \\
		\ket{b}\qquad & \targ    & ~~\qquad\qquad\quad\ket{a}\ket{b} \oplus \ket{a}\qw \\
		}
  
  }
	\caption{CX Gate representation for $\ket{a}\ket{b}~=~\ket{a}\ket{b}~\oplus~\ket{a}$\label{fig:cxgate}}
\end{figure}
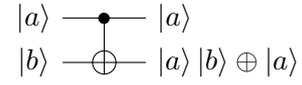

Moving on, we have the CX gate, also known as the CNOT gate, which is a two-qubit gate having one control and one target qubit. It performs a NOT operation on the target qubit if the control qubit is in the state $\ket{1}$, and is given in Fig.~\ref{fig:cxgate}. Here, the black dot within the CNOT gate signifies that when the controlling bit (represented by the black dot) holds the value \emph{1}, the NOT gate on the target bit is activated. Conversely, the white dot indicates that when the controlling bit is \emph{0} the NOT gate operates on the target bit. Both the NOT and the Controlled NOT are fundamental gates for having a quantum cost and delay of 1  \cite{xia2018efficient}. We use the definition of quantum cost as the number of fundamental unitary and binary reversible gates (e.g. NOT and CNOT) that are used in the design of a gate in a general decomposition sense \cite{thapliyal2010design}. This is then extended to quantum delay, defined as the measure of the logical depth of a circuit, such that the delay of fundamental unitary and binary reversible gates would be 1$\Delta$ \cite{thapliyal2010design}.

\begin{figure}
	\centering
		\scalebox{1}{

        \Qcircuit @C=1em @R=1em {
             \lstick{\ket{a}} & \ctrl{1} & \rstick{\ket{a}} \qw \\
                    \lstick{\ket{b}} & \gate{V+} & \rstick{V^+\ket{b}} \qw \\
            }
                }
	\caption{Controlled $V^+$ Gate representation for $\ket{a}\ket{b} \mapsto V^+(\ket{b})$ (unitary transformation)\label{fig:Vplusgate}}
\end{figure}

\begin{figure}
	\centering
		\scalebox{1}{

        \Qcircuit @C=1em @R=1em {
             \lstick{\ket{a}} & \ctrl{1} & \rstick{\ket{a}} \qw \\
                    \lstick{\ket{b}} & \gate{V-} & \rstick{V^-\ket{b}} \qw \\
            }
                }
	\caption{Controlled V Gate representation for $\ket{a}\ket{b} \mapsto V^-(\ket{b})$ (unitary transformation)\label{fig:Vminusgate}}
\end{figure}

V and $V^+$ represent two useful quantum gates with a quantum cost and delay each equivalent to 1 \cite{thapliyal2010design, xia2018efficient}. These gates are defined by their respective unitary matrices, which adhere to the mathematical property \(VV^+ = V^+V = 1\), signifying their unitarity and the inherent reversibility of quantum operations. Geometrically, these gates are interpreted as a quarter turn about the $x$-axis in a Bloch sphere. These matrices are precisely defined as:
\begin{equation}
V = \frac{i + 1}{2} \begin{pmatrix} 1 & -i \\ -i & 1 \end{pmatrix}    
\end{equation}
\begin{equation}
V^+ = \frac{1 - i}{2} \begin{pmatrix} 1 & i \\ i & 1 \end{pmatrix}    
\end{equation}
where \(i\) represents the imaginary unit.

A ternary case is that of the CCX gate; also known as the Toffoli gate. It operates on $3$-qubits by performing a bit-wise logical AND on the first two qubits and then applies its result to the third qubit as an XOR, resulting in a flip in $\ket{c}$, provided that both the first and second qubits are in state $\ket{1}$. This is illustrated in Fig.~\ref{fig:ccxgate}. The quantum cost of a Toffoli gate is determined by considering the number of fundamental quantum gates it comprises, such as CNOT gates, controlled-V gates, and controlled-$V^+$ gates. In the case of a Toffoli gate, it involves two CNOT gates, two controlled-V gates, and one controlled-$V^+$ gate. Therefore, both the quantum cost and delay of a Toffoli gate are evaluated at 5$\triangle$   \cite{thapliyal2010design, xia2018efficient}.
The CCX can be realized through the use of XOR and AND operations. Considering the case of $\ket{a}$, $\ket{b}$, and $\ket{c}$ (Figure~\ref{fig:ccxgate}), then CCX can be achieved as $\ket{a}\ket{b}\ket{c \oplus ab}$. This is realized in the first step through an AND $\mathbf{x} = \mathbf{a_2} \land \mathbf{b_2}$. The intermediate $\mathbf{x}$ is then passed as an XOR to get the final result $\mathbf{x'} =  \mathbf{c} \oplus \mathbf{x}$.

\begin{figure}
	\centering
		\scalebox{1}{
		\Qcircuit @C=1em @R=1em {
			\ket{a}\qquad & \ctrl{1} & \qquad\ket{a}\qw  \\
			\ket{b}\qquad & \ctrl{1} & \qquad\ket{b}\qw  \\
			\ket{c}\qquad & \targ & \quad\qquad\qquad\qquad\ket{a}\ket{b}\ket{c \oplus a b}\qw  \\
		}
	}
	\caption{CCX Gate representation for $\ket{a}\ket{b}\ket{c}~=~\ket{a}\ket{b}\ket{c \oplus a b}$\label{fig:ccxgate}}
\end{figure}

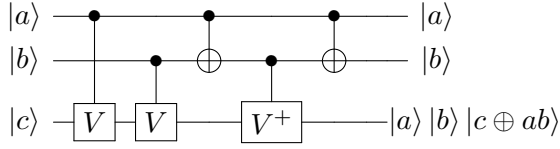
\begin{figure}
\centering
\scalebox{1}{
	\Qcircuit @C=0.7em @R=1em {\\
    \ket{a}  \qquad  & \ctrl{2} & \qw & \ctrl{1} & \qw & \ctrl{1} & \qw & \qw & \qquad\ket{a}\qw  \\
    \ket{b}  \qquad   & \qw & \ctrl{1} & \targ & \ctrl{1} & \targ & \qw & \qw & \qquad\ket{b}\qw \\
    \ket{c} \qquad   &  \gate{V}  &  \gate{V}  & \qw &  \gate{V^+}  & \qw & \qw & \qquad \qquad  \qquad\ket{a}\ket{b}\ket{c \oplus a b}\qw\\
\\ }
}
\caption{CCX Gate Representation using V and V+ gate}
\label{fig:circuit}
\end{figure}

In the next sections, we discuss techniques that implement the GQBSC using these gates. 

\section{Related Work}

\begin{figure}
	\centering
	\includegraphics[width=0.5\textwidth]{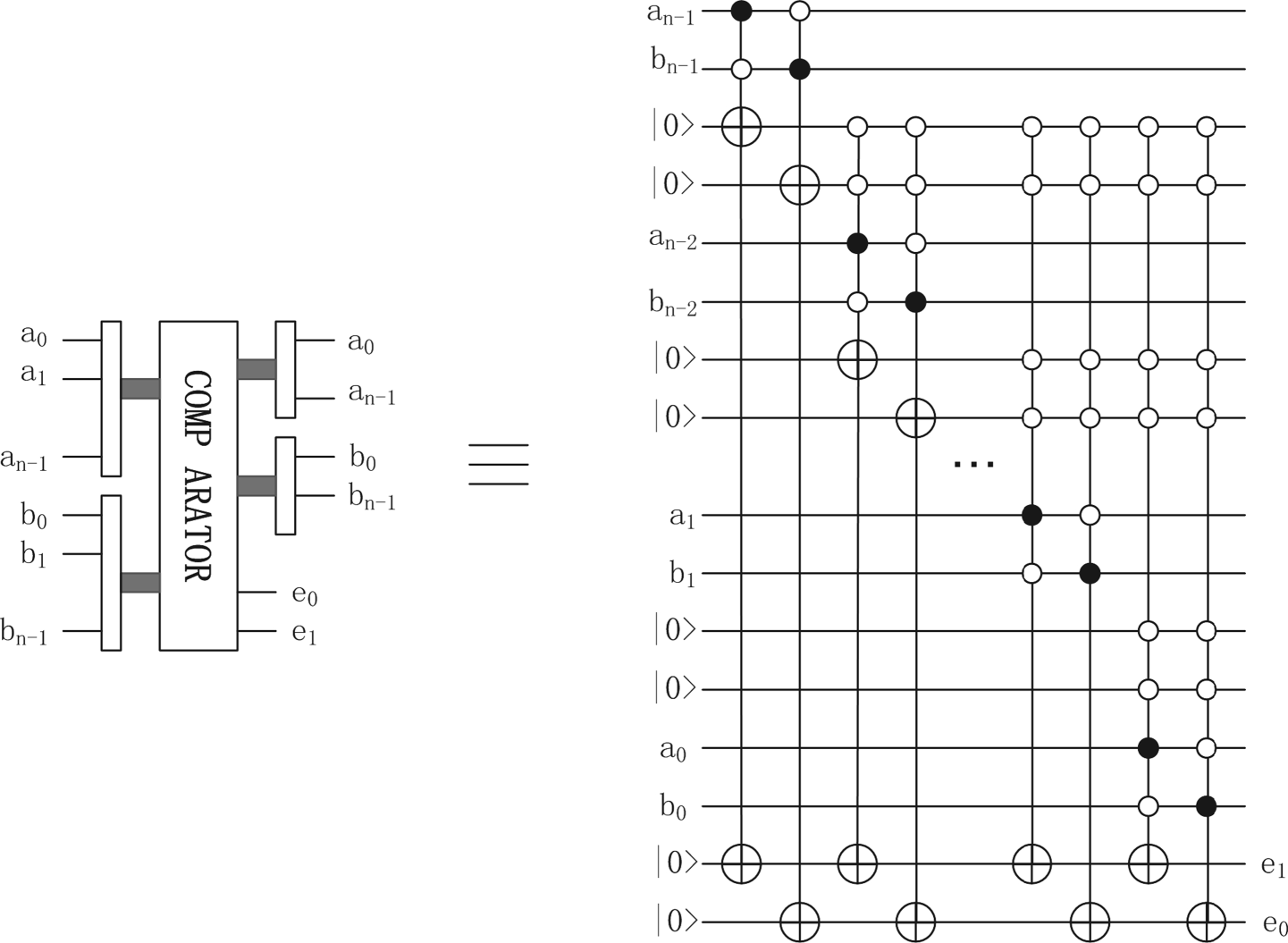} 
	\caption{The reversible comparator presented by Want et al.  \cite{wang2012design}.}
	\label{fig:a}
\end{figure}

\begin{figure}
	\centering
	\includegraphics[width=0.45\textwidth]{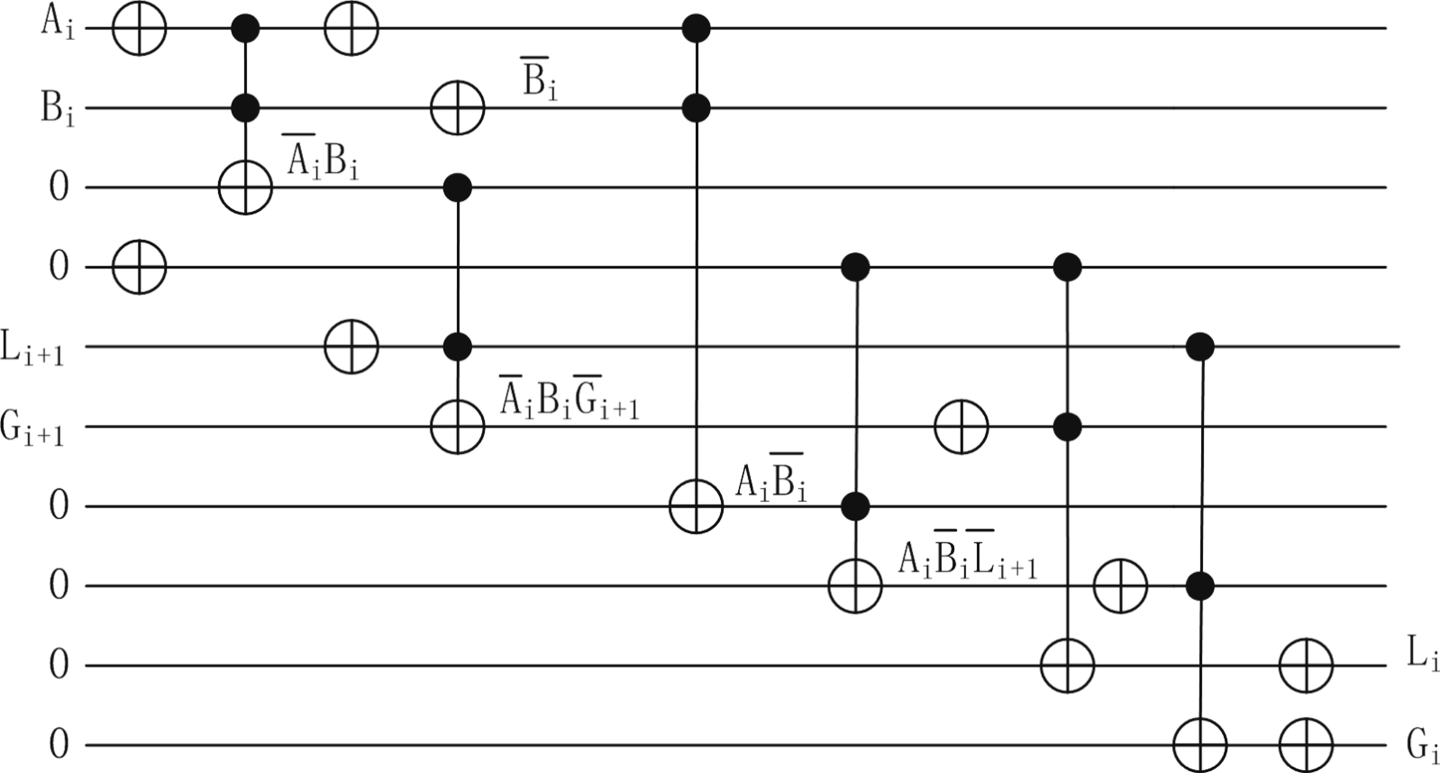} 
	\caption{The comparator presented by Al-Rabadi et al. \cite{al2009closed}.}
	\label{fig:b}
\end{figure}

\begin{figure}
	\centering
	\includegraphics[width=0.5\textwidth]{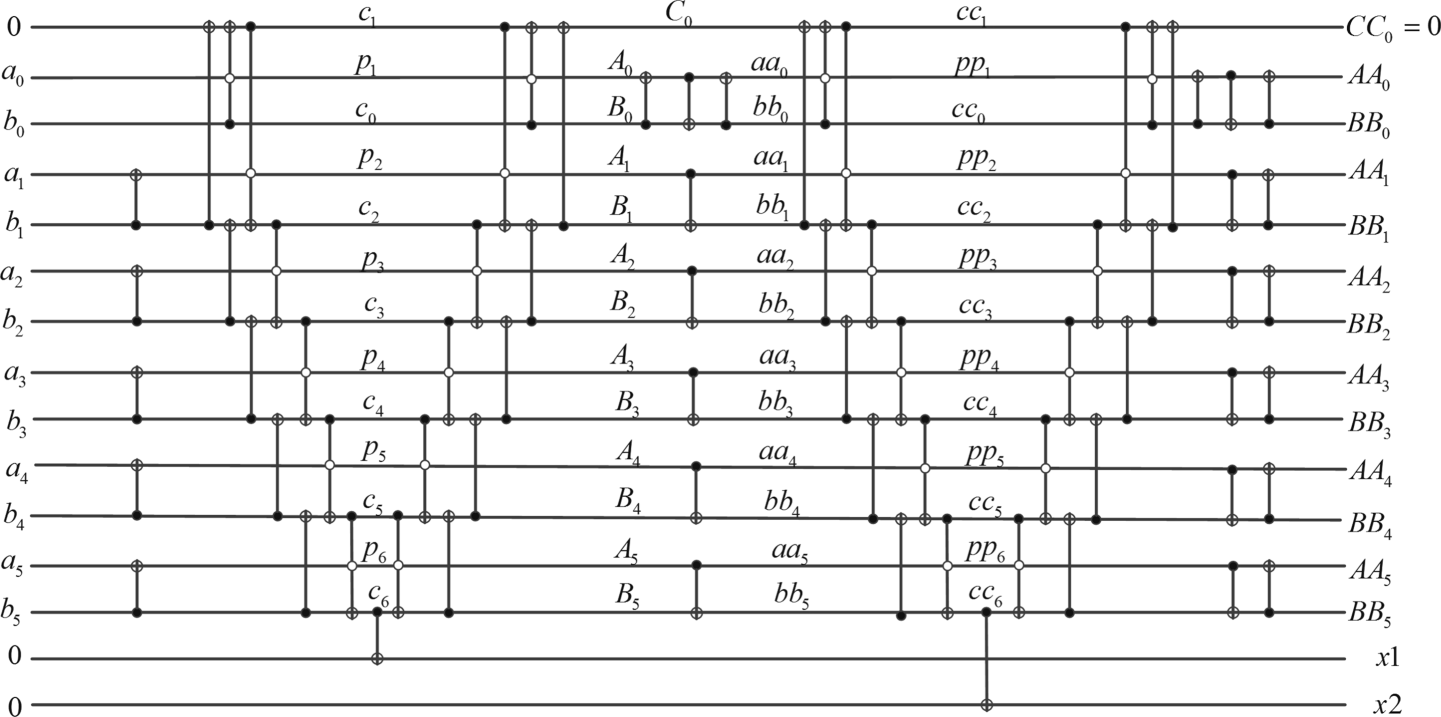} 
	\caption{The comparator presented by Haiying Xia et al.\cite{xia2018efficient}.}
	\label{fig:d}
\end{figure}

\begin{figure}
	\centering
	\includegraphics[width=0.5\textwidth]{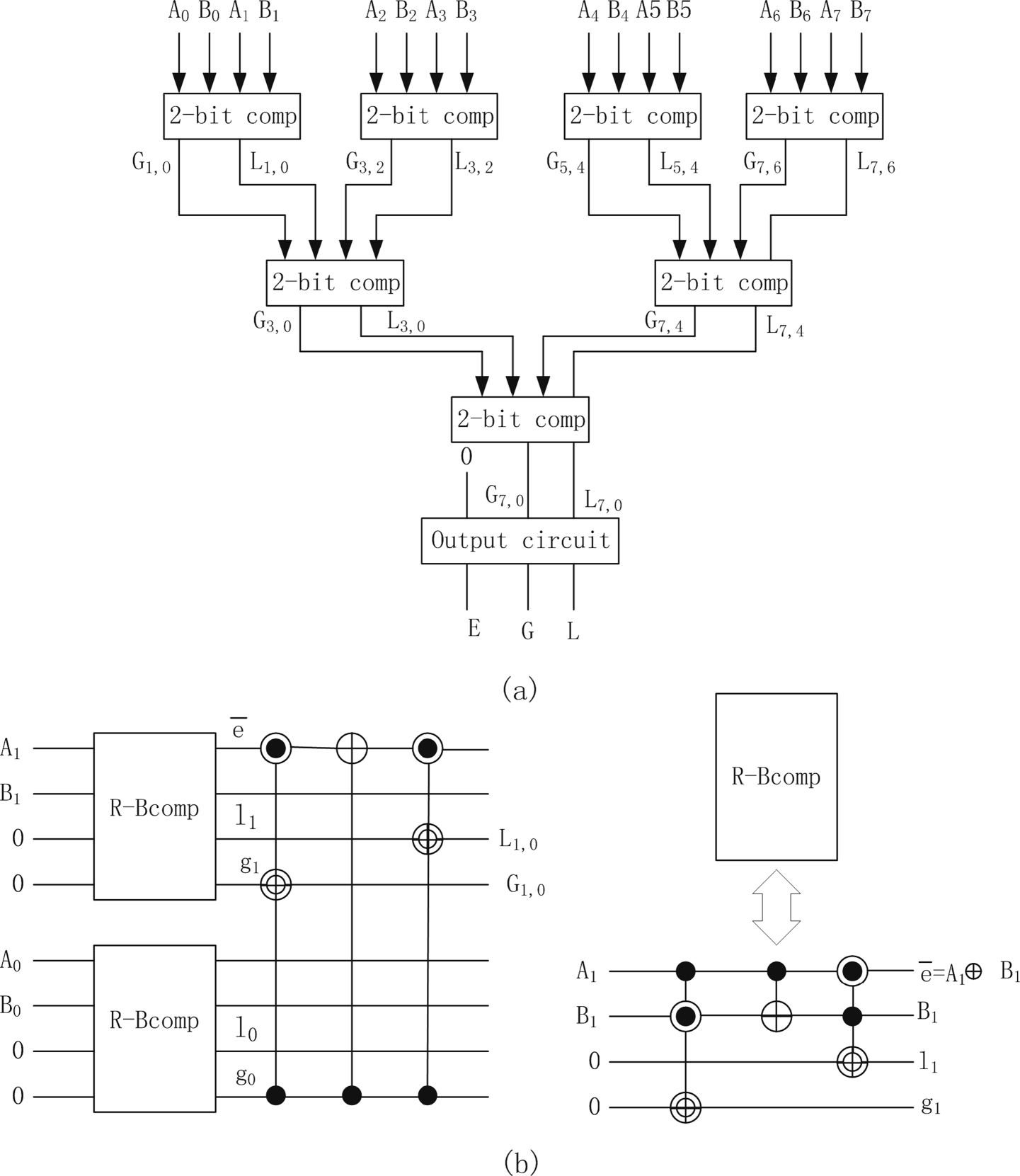} 
	\caption{There are (a) and (b) the tree-based comparator presented by Thapliyal et al. \cite{thapliyal2010design}.}
	\label{fig:c}
\end{figure}

There have been various quantum comparators reported in the literature. Wang et al. introduced a quantum comparator that accomplishes the comparison of two quantum logic states, each having $n$ quantum bits, in a sequential manner \cite{wang2012design}. The circuit involves the use of $2n-2$ ancillary input bits, as depicted in (Figure \ref{fig:a}). In this diagram, $|e_0\rangle$ and $|e_1\rangle$ represent the meaningful outcomes.
Al-Rabadi proposed a sequentially structured quantum comparator that links together a sequence of 1-bit comparators, as shown in (Figure \ref{fig:b}). This arrangement requires 6 ancillary input bits for each 1-bit comparator \cite{al2009closed}.
Thapliyal et al. devised a tree-based comparator, illustrated in (Figure \ref{fig:c}), wherein every node corresponds to a 2-bit comparator demonstrated in (Figure \ref{fig:c}). This specialized comparator can assess 2-bit binary numbers \cite{thapliyal2010design}.
Vudadha et al. enhanced the tree-based comparator using a prefix tree \cite{vudadha2012design}. This design comprises three stages: the initial stage integrates a 1-bit comparator featuring two meaningful outputs. The outputs of the 1-bit comparator phase are then grouped in the second stage using prefix grouping to generate the final outputs $G$. While the tree-based quantum comparator surpasses the sequential-based comparator in terms of time delay, it lags behind the sequential-based comparator in the number of ancillary bits required.
Xia et al. proposed a new serial base comparator (Figure \ref{fig:d}) that can compare two n-size numbers using one ancillary qubit, but the quantum cost and delay are still high \cite{xia2018efficient}.

\begin{figure}
	\centering
	\includegraphics[width=0.5\textwidth]{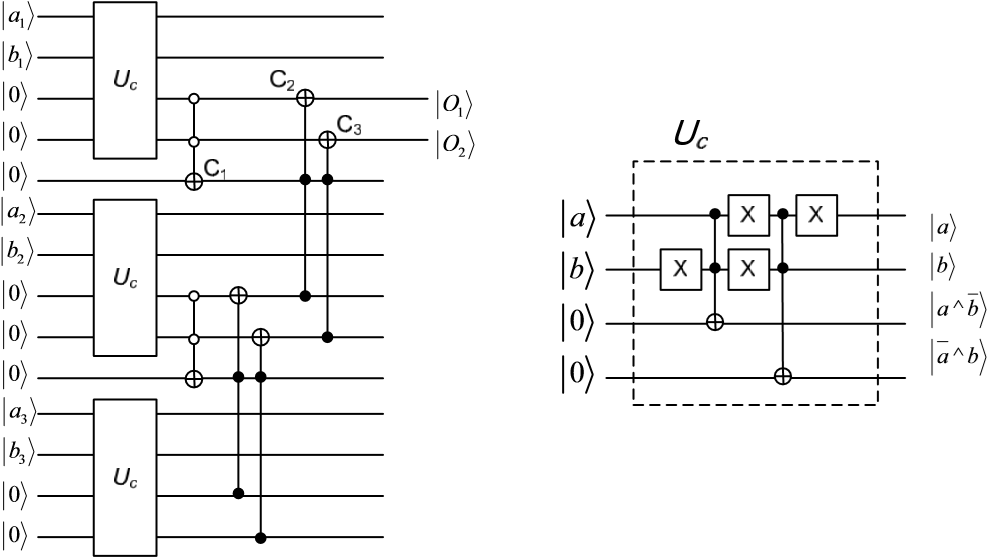} 
	\caption{The comparator presented by David et al.\cite{oliveira2007quantum}.}
	\label{fig:e}
\end{figure}

\section{Proposed GQBSC}

Our objective is to conduct a comparison between two n-bit numbers, denoted as \emph{a} and \emph{b}, utilizing only two ancillary bits. We employ two binary strings of length $n$ such that \emph{a} is represented as $a_{n-1} \ldots a_0$, where $a_0$ is the most significant bit. Similarly, \emph{b} is represented as $b_{n-1} \ldots b_0$, where $b_0$ is the most significant bit. In case the strings are unequal, padding is performed on the smaller string up to the length of the larger string. To characterize the outputs of  $a_i$, and $b_i$, we now present a discussion of our proposed approach, described in Algo.~\ref{algo:GeneralizeAlgorithm}. 

The approach takes as input the vectors $\ket{a}$, $\ket{b}$, and registers $\ket{r_0}$, $\ket{r_1}$ initialized to $\ket{0}$. The algorithm returns $\ket{r_0}$, $\ket{r_1}$ as classical variables $c_0$ and $c_1$. The algorithm works by performing a 1-bit quantum bit string comparator (Algo.~\ref{equation:1BitC}) on the most significant bit $a_0$ and $b_0$ and collects its intermediate output in $r_0$ and $r_1$. For each of the remaining bit string lengths, the intermediate outputs are approximated against $\ket{0}$, and if true, the 1-bit GQBSC is carried out for the next significant bit. However, if $r_1$ is approximated to $\ket{1}$, a bit flip is carried out on $r_0$ (through Pauli X gate). 

The 1-bit GQBSC (Algo.~\ref{equation:1BitC}) is the realization of the quantum circuit given in Fig.~\ref{fig:1bit}. It is essentially a core component of our proposed model and utilizes four qubits; $\ket{a}$ and $\ket{b}$ for storing the two input bitstrings, and $\ket{r_0}$ and $\ket{r_1}$ for storing intermediate comparisons. The comparison is itself carried out through the successive application of multiple NOT and CCX gates as shown. The intermediate states are then stored in classical registers $cr$ through $Z$-measurement gates. The output of the proposed circuit exhibits four possible binary states; \emph{00} indicating $a \approx b$, \emph{10} indicating $a > b$, and \emph{01} and \emph{11} both indicating $a < b$. 

\begin{algorithm}
	\caption{ Generalized Quantum Bit String Comparator(GQBSC)}
	\SetAlgoLined
	
	\KwData{$\ket{a}$, $\ket{b}$, $\ket{r_0} = \ket{r_1} = \ket{0}$}
	\KwResult{$c_0$, $c_1$}
    $[r_0, r_1] \gets \text{1BITGQBSC}(a_0, b_0, r_0, r_1)$\\
	\For{$i$ in $1 \ldots n-1$}{
	   \If{ $(r_0 \approx \ket{0}) \text{ and } (r_1 \approx \ket{0})$ }{
		$[r_0, r_1] \gets \text{1BITGQBSC}(a_i, b_i, r_0, r_1)$ \\
	   }
	   \If{ $(r_1 \approx \ket{1})$}{
		$r_0 \gets \mathrm{X} r_0$ \\
	   }
	}
	$[c_0, c_1] \gets [r_0, r_1]$\\~\\
 \label{algo:GeneralizeAlgorithm}
\end{algorithm}

\begin{algorithm}
	\caption{One bit Quantum Bit String Comparator (1BITGQBSC)\label{equation:1BitC}}
	\SetAlgoLined
    \SetKwProg{Fn}{1BITGQBSC}{}{end}
	\KwResult{$r_0$, $r_1$}
    \Fn{$(\ket{a},\ket{b},r_0,r_1)$}{
        $b \gets \mathrm{X}b$\\
        $r_0 \gets r_0 \oplus (a_1 \land b_1)$ \\
		$a \gets \mathrm{X} a$ \\
		$b \gets \mathrm{X} b$ \\
		$r1 \gets r1 \oplus (a_1 \land b_1)$ \\
		$a \gets \mathrm{X} a$
    }
\end{algorithm}
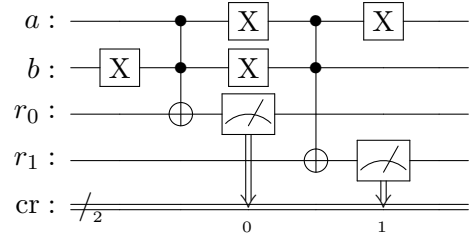
\begin{figure}
	\centering
	\scalebox{1.0}{
		\Qcircuit @C=1.0em @R=0.2em @!R { \\
			\nghost{{a} :  } & \lstick{{a} :  } & \qw & \ctrl{1} & \gate{\mathrm{X}} & \ctrl{1} & \gate{\mathrm{X}} & \qw & \qw\\
			\nghost{{b} :  } & \lstick{{b} :  } & \gate{\mathrm{X}} & \ctrl{1} & \gate{\mathrm{X}} & \ctrl{2} & \qw & \qw & \qw\\
			\nghost{{r}_{0} :  } & \lstick{{r}_{0} :  } & \qw & \targ & \meter & \qw & \qw & \qw & \qw\\
			\nghost{{r}_{1} :  } & \lstick{{r}_{1} :  } & \qw & \qw & \qw & \targ & \meter & \qw & \qw\\
			\nghost{\mathrm{{cr} :  }} & \lstick{\mathrm{{cr} :  }} & \lstick{/_{_{2}}} \cw & \cw & \dstick{_{_{\hspace{0.0em}0}}} \cw \ar @{<=} [-2,0] & \cw & \dstick{_{_{\hspace{0.0em}1}}} \cw \ar @{<=} [-1,0] & \cw & \cw\\
			\\ }}
	\caption{Generalized Quantum Bit String Comparator (GQBSC) Circuit for two inputs with a maximum size of 1 bit}
	\label{fig:1bit}
\end{figure}

\begin{figure*}
\centering
\scalebox{1}{
\Qcircuit @C=0.8em @R=0.09em @!R { \\
	 		 	\nghost{{a}_{0} :  } & \lstick{{a}_{0} :  } & \gate{\mathrm{X}} & \multigate{7}{\mathrm{1BC}}_<<<{0} & \qw & \qw & \qw & \qw & \qw & \qw & \qw & \qw & \qw & \qw & \qw\\
	 	\nghost{{a}_{1} :  } & \lstick{{a}_{1} :  } & \gate{\mathrm{X}} & \ghost{\mathrm{1BC}} & \qw & \qw & \multigate{6}{\mathrm{1BC}}_<<<{0} & \qw & \qw & \qw & \qw & \qw & \qw & \qw & \qw\\
	 	\nghost{{a}_{2} :  } & \lstick{{a}_{2} :  } & \gate{\mathrm{X}} & \ghost{\mathrm{1BC}} & \qw & \qw & \ghost{\mathrm{1BC}} & \qw & \qw & \qw & \multigate{5}{\mathrm{1BC}}_<<<{0} & \qw & \qw & \qw & \qw\\
	 	\nghost{{b}_{0} :  } & \lstick{{b}_{0} :  } & \qw & \ghost{\mathrm{1BC}}_<<<{1} & \qw & \qw & \ghost{\mathrm{1BC}} & \qw & \qw & \qw & \ghost{\mathrm{1BC}} & \qw & \qw & \qw & \qw\\
	 	\nghost{{b}_{1} :  } & \lstick{{b}_{1} :  } & \gate{\mathrm{X}} & \ghost{\mathrm{1BC}} & \qw & \qw & \ghost{\mathrm{1BC}}_<<<{1} & \qw & \qw & \qw & \ghost{\mathrm{1BC}} & \qw & \qw & \qw & \qw\\
	 	\nghost{{b}_{2} :  } & \lstick{{b}_{2} :  } & \gate{\mathrm{X}} & \ghost{\mathrm{1BC}} & \qw & \qw & \ghost{\mathrm{1BC}} & \qw & \qw & \qw & \ghost{\mathrm{1BC}}_<<<{1} & \qw & \qw & \qw & \qw\\
	 	\nghost{{r}_{0} :  } & \lstick{{r}_{0} :  } & \qw & \ghost{\mathrm{1BC}}_<<<{2} & \meter & \qw & \ghost{\mathrm{1BC}}_<<<{2} & \meter & \qw & \gate{\mathrm{X}} & \ghost{\mathrm{1BC}}_<<<{2} & \meter & \qw & \qw & \qw\\
	 	\nghost{{r}_{1} :  } & \lstick{{r}_{1} :  } & \qw & \ghost{\mathrm{1BC}}_<<<{3} & \qw & \meter & \ghost{\mathrm{1BC}}_<<<{3} & \qw & \meter & \qw & \ghost{\mathrm{1BC}}_<<<{3} & \qw & \meter & \qw & \qw\\
	 	\nghost{\mathrm{{cr} :  }} & \lstick{\mathrm{{cr} :  }} & \lstick{/_{_{2}}} \cw & \controlo \cw^(0.0){^{\mathtt{0x0}}} \cwx[-1] & \dstick{_{_{\hspace{0.0em}0}}} \cw \ar @{<=} [-2,0] & \dstick{_{_{\hspace{0.0em}1}}} \cw \ar @{<=} [-1,0] & \controlo \cw^(0.0){^{\mathtt{0x0}}} \cwx[-1] & \dstick{_{_{\hspace{0.0em}0}}} \cw \ar @{<=} [-2,0] & \dstick{_{_{\hspace{0.0em}1}}} \cw \ar @{<=} [-1,0] & \control \cw^(0.0){^{\mathtt{0x2}}} \cwx[-2] & \controlo \cw^(0.0){^{\mathtt{0x0}}} \cwx[-1] & \dstick{_{_{\hspace{0.0em}0}}} \cw \ar @{<=} [-2,0] & \dstick{_{_{\hspace{0.0em}1}}} \cw \ar @{<=} [-1,0] & \cw & \cw\\
\\ }}

\caption{Generalized Quantum Bit String Comparator(GQBSC) Circuit for 3-bit input size.}
\label{fig:3bit}
\end{figure*}
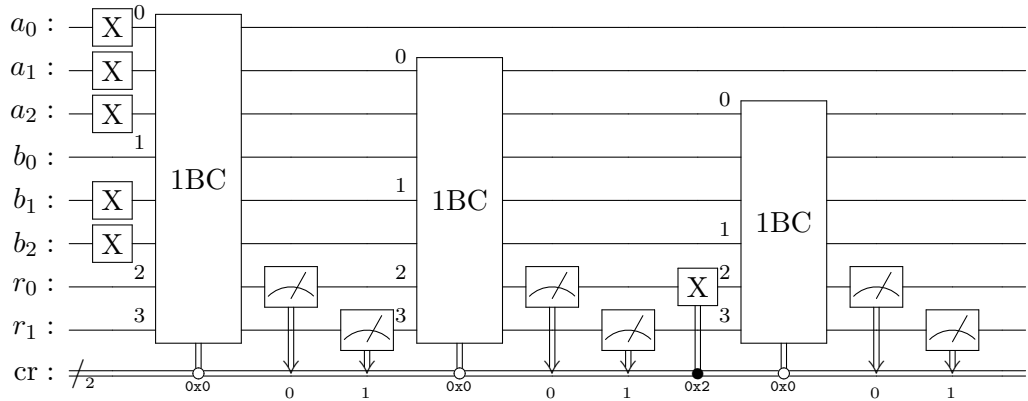

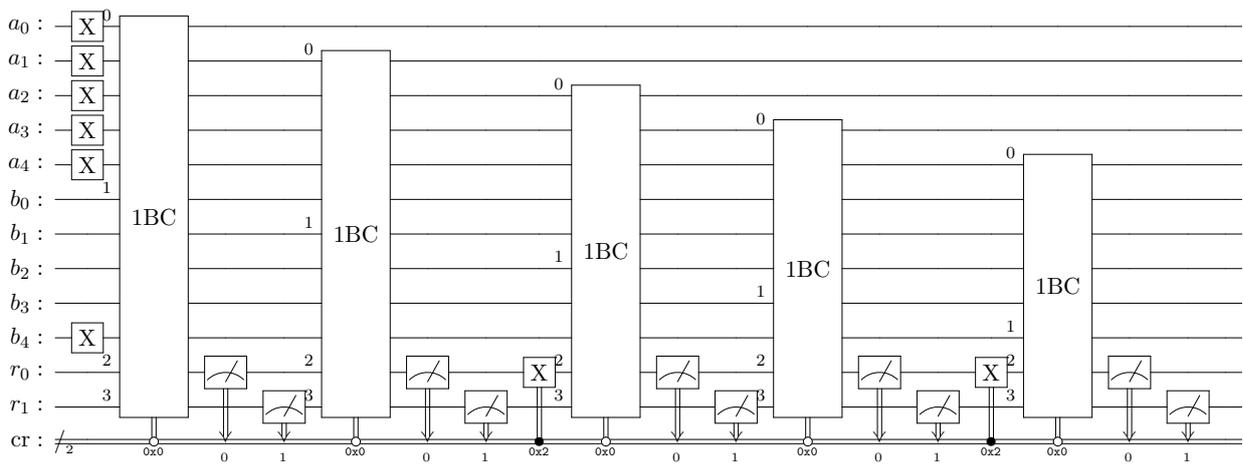
\begin{figure*}
\centering
\scalebox{.8}{
\Qcircuit @C=0.7em @R=0.09em @!R { \\
	 	\nghost{{a}_{0} :  } & \lstick{{a}_{0} :  } & \gate{\mathrm{X}} & \multigate{11}{\mathrm{1BC}}_<<<{0} & \qw & \qw & \qw & \qw & \qw & \qw & \qw & \qw & \qw & \qw & \qw & \qw & \qw & \qw & \qw & \qw & \qw & \qw\\
	 	\nghost{{a}_{1} :  } & \lstick{{a}_{1} :  } & \gate{\mathrm{X}} & \ghost{\mathrm{1BC}} & \qw & \qw & \multigate{10}{\mathrm{1BC}}_<<<{0} & \qw & \qw & \qw & \qw & \qw & \qw & \qw & \qw & \qw & \qw & \qw & \qw & \qw & \qw & \qw\\
	 	\nghost{{a}_{2} :  } & \lstick{{a}_{2} :  } & \gate{\mathrm{X}} & \ghost{\mathrm{1BC}} & \qw & \qw & \ghost{\mathrm{1BC}} & \qw & \qw & \qw & \multigate{9}{\mathrm{1BC}}_<<<{0} & \qw & \qw & \qw & \qw & \qw & \qw & \qw & \qw & \qw & \qw & \qw\\
	 	\nghost{{a}_{3} :  } & \lstick{{a}_{3} :  } & \gate{\mathrm{X}} & \ghost{\mathrm{1BC}} & \qw & \qw & \ghost{\mathrm{1BC}} & \qw & \qw & \qw & \ghost{\mathrm{1BC}} & \qw & \qw & \multigate{8}{\mathrm{1BC}}_<<<{0} & \qw & \qw & \qw & \qw & \qw & \qw & \qw & \qw\\
	 	\nghost{{a}_{4} :  } & \lstick{{a}_{4} :  } & \gate{\mathrm{X}} & \ghost{\mathrm{1BC}} & \qw & \qw & \ghost{\mathrm{1BC}} & \qw & \qw & \qw & \ghost{\mathrm{1BC}} & \qw & \qw & \ghost{\mathrm{1BC}} & \qw & \qw & \qw & \multigate{7}{\mathrm{1BC}}_<<<{0} & \qw & \qw & \qw & \qw\\
	 	\nghost{{b}_{0} :  } & \lstick{{b}_{0} :  } & \qw & \ghost{\mathrm{1BC}}_<<<{1} & \qw & \qw & \ghost{\mathrm{1BC}} & \qw & \qw & \qw & \ghost{\mathrm{1BC}} & \qw & \qw & \ghost{\mathrm{1BC}} & \qw & \qw & \qw & \ghost{\mathrm{1BC}} & \qw & \qw & \qw & \qw\\
	 	\nghost{{b}_{1} :  } & \lstick{{b}_{1} :  } & \qw & \ghost{\mathrm{1BC}} & \qw & \qw & \ghost{\mathrm{1BC}}_<<<{1} & \qw & \qw & \qw & \ghost{\mathrm{1BC}} & \qw & \qw & \ghost{\mathrm{1BC}} & \qw & \qw & \qw & \ghost{\mathrm{1BC}} & \qw & \qw & \qw & \qw\\
	 	\nghost{{b}_{2} :  } & \lstick{{b}_{2} :  } & \qw & \ghost{\mathrm{1BC}} & \qw & \qw & \ghost{\mathrm{1BC}} & \qw & \qw & \qw & \ghost{\mathrm{1BC}}_<<<{1} & \qw & \qw & \ghost{\mathrm{1BC}} & \qw & \qw & \qw & \ghost{\mathrm{1BC}} & \qw & \qw & \qw & \qw\\
	 	\nghost{{b}_{3} :  } & \lstick{{b}_{3} :  } & \qw & \ghost{\mathrm{1BC}} & \qw & \qw & \ghost{\mathrm{1BC}} & \qw & \qw & \qw & \ghost{\mathrm{1BC}} & \qw & \qw & \ghost{\mathrm{1BC}}_<<<{1} & \qw & \qw & \qw & \ghost{\mathrm{1BC}} & \qw & \qw & \qw & \qw\\
	 	\nghost{{b}_{4} :  } & \lstick{{b}_{4} :  } & \gate{\mathrm{X}} & \ghost{\mathrm{1BC}} & \qw & \qw & \ghost{\mathrm{1BC}} & \qw & \qw & \qw & \ghost{\mathrm{1BC}} & \qw & \qw & \ghost{\mathrm{1BC}} & \qw & \qw & \qw & \ghost{\mathrm{1BC}}_<<<{1} & \qw & \qw & \qw & \qw\\
	 	\nghost{{r}_{0} :  } & \lstick{{r}_{0} :  } & \qw & \ghost{\mathrm{1BC}}_<<<{2} & \meter & \qw & \ghost{\mathrm{1BC}}_<<<{2} & \meter & \qw & \gate{\mathrm{X}} & \ghost{\mathrm{1BC}}_<<<{2} & \meter & \qw & \ghost{\mathrm{1BC}}_<<<{2} & \meter & \qw & \gate{\mathrm{X}} & \ghost{\mathrm{1BC}}_<<<{2} & \meter & \qw & \qw & \qw\\
	 	\nghost{{r}_{1} :  } & \lstick{{r}_{1} :  } & \qw & \ghost{\mathrm{1BC}}_<<<{3} & \qw & \meter & \ghost{\mathrm{1BC}}_<<<{3} & \qw & \meter & \qw & \ghost{\mathrm{1BC}}_<<<{3} & \qw & \meter & \ghost{\mathrm{1BC}}_<<<{3} & \qw & \meter & \qw & \ghost{\mathrm{1BC}}_<<<{3} & \qw & \meter & \qw & \qw\\
	 	\nghost{\mathrm{{cr} :  }} & \lstick{\mathrm{{cr} :  }} & \lstick{/_{_{2}}} \cw & \controlo \cw^(0.0){^{\mathtt{0x0}}} \cwx[-1] & \dstick{_{_{\hspace{0.0em}0}}} \cw \ar @{<=} [-2,0] & \dstick{_{_{\hspace{0.0em}1}}} \cw \ar @{<=} [-1,0] & \controlo \cw^(0.0){^{\mathtt{0x0}}} \cwx[-1] & \dstick{_{_{\hspace{0.0em}0}}} \cw \ar @{<=} [-2,0] & \dstick{_{_{\hspace{0.0em}1}}} \cw \ar @{<=} [-1,0] & \control \cw^(0.0){^{\mathtt{0x2}}} \cwx[-2] & \controlo \cw^(0.0){^{\mathtt{0x0}}} \cwx[-1] & \dstick{_{_{\hspace{0.0em}0}}} \cw \ar @{<=} [-2,0] & \dstick{_{_{\hspace{0.0em}1}}} \cw \ar @{<=} [-1,0] & \controlo \cw^(0.0){^{\mathtt{0x0}}} \cwx[-1] & \dstick{_{_{\hspace{0.0em}0}}} \cw \ar @{<=} [-2,0] & \dstick{_{_{\hspace{0.0em}1}}} \cw \ar @{<=} [-1,0] & \control \cw^(0.0){^{\mathtt{0x2}}} \cwx[-2] & \controlo \cw^(0.0){^{\mathtt{0x0}}} \cwx[-1] & \dstick{_{_{\hspace{0.0em}0}}} \cw \ar @{<=} [-2,0] & \dstick{_{_{\hspace{0.0em}1}}} \cw \ar @{<=} [-1,0] & \cw & \cw\\
\\ }}

\caption{Generalized Quantum Bit String Comparator(GQBSC) Circuit for 5-bit input size.}
\label{fig:5bit}
\end{figure*}

The above is a description of the formation of a generalized GQBSC circuit. The mean circuit size grows and shrinks on the basis of input lengths. While Fig.~\ref{fig:1bit} illustrates a circuit for 1-bit GQBSC requiring 4 qubits, Fig.~\ref{fig:2bit} shows the pattern by which it grows for a two-bit size, having a requirement of six qubits. Here, $a_0$ and $a_1$ are used for storing the information of the first number, while $b_0$ and $b_1$ are used for storing the second number. The 1-bit Comparator is successively applied to Qubits $a_0$ and $b_0$, and then to $a_1$ and $b_1$ after $Z$-measurements. Similar representation are provided for 3-bit (Fig.~\ref{fig:3bit}) and 5-bit (Fig.~\ref{fig:5bit}) inputs. 

\begin{figure}
\scalebox{0.6}{
\Qcircuit @C=0.8em @R=0.09em @!R { \\
	 	\nghost{{a}_{0} :  } & \lstick{{a}_{0} :  } & \gate{\mathrm{X}} & \multigate{5}{\mathrm{1BC}}_<<<{0} & \qw & \qw & \qw & \qw & \qw & \qw & \qw & \qw & \qw & \qw\\
\nghost{{a}_{1} :  } & \lstick{{a}_{1} :  } & \qw & \ghost{\mathrm{1BC}} & \qw & \qw & \multigate{4}{\mathrm{1BC}}_<<<{0} & \qw & \qw & \qw & \qw & \qw & \qw & \qw\\
\nghost{{b}_{0} :  } & \lstick{{b}_{0} :  } & \gate{\mathrm{X}} & \ghost{\mathrm{1BC}}_<<<{1} & \qw & \qw & \ghost{\mathrm{1BC}} & \qw & \qw & \qw & \qw & \qw & \qw & \qw\\
\nghost{{b}_{1} :  } & \lstick{{b}_{1} :  } & \gate{\mathrm{X}} & \ghost{\mathrm{1BC}} & \qw & \qw & \ghost{\mathrm{1BC}}_<<<{1} & \qw & \qw & \qw & \qw & \qw & \qw & \qw\\
\nghost{{r}_{0} :  } & \lstick{{r}_{0} :  } & \qw & \ghost{\mathrm{1BC}}_<<<{2} & \meter & \qw & \ghost{\mathrm{1BC}}_<<<{2} & \meter & \qw & \gate{\mathrm{X}} & \meter & \qw & \qw & \qw\\
\nghost{{r}_{1} :  } & \lstick{{r}_{1} :  } & \qw & \ghost{\mathrm{1BC}}_<<<{3} & \qw & \meter & \ghost{\mathrm{1BC}}_<<<{3} & \qw & \meter & \qw & \qw & \meter & \qw & \qw\\
\nghost{\mathrm{{cr} :  }} & \lstick{\mathrm{{cr} :  }} & \lstick{/_{_{2}}} \cw & \controlo \cw^(0.0){^{\mathtt{0x0}}} \cwx[-1] & \dstick{_{_{\hspace{0.0em}0}}} \cw \ar @{<=} [-2,0] & \dstick{_{_{\hspace{0.0em}1}}} \cw \ar @{<=} [-1,0] & \controlo \cw^(0.0){^{\mathtt{0x0}}} \cwx[-1] & \dstick{_{_{\hspace{0.0em}0}}} \cw \ar @{<=} [-2,0] & \dstick{_{_{\hspace{0.0em}1}}} \cw \ar @{<=} [-1,0] & \control \cw^(0.0){^{\mathtt{0x2}}} \cwx[-2] & \dstick{_{_{\hspace{0.0em}0}}} \cw \ar @{<=} [-2,0] & \dstick{_{_{\hspace{0.0em}1}}} \cw \ar @{<=} [-1,0] & \cw & \cw\\
\\ }}
\caption{Generalized Quantum Bit String Comparator (GQBSC) Circuit for two-bit size.}
\label{fig:2bit}
\end{figure}
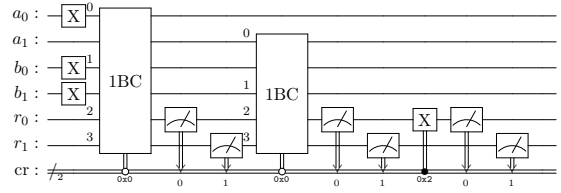

\section{Results}
Quantum computing relies on precise measurements for computation outcomes. Traditionally, measurements were only done at the end to prevent errors. IBM's new hardware-based approach allows dynamic circuits and mid-computation measurements, offering three key benefits;  reduced qubit usage, fewer additional operations, and improved accuracy 
 \cite{hua2023caqr}.

Given our model, we now present a discussion on its validation, verification, and behavior, all of which are tested with respect to different numbers of input sizes. The Verification of each test case is carried out from single-bit up to 1000-bits; a portion of which is illustrated in Table.~\ref{tab:verification}. 

\begin{table*}
\begin{adjustwidth}{-1in}{-1in} 
\setlength{\tabcolsep}{7pt} 
    \centering
    \begin{tabular}{|c|c|c|c|c|c|c|c|c|}
        \hline
        \multirow{2}{*}{Sr.}&\multirow{2}{*}{$(\ket{a})$} &\multirow{2}{*} {bin $(\ket{a})$} &\multirow{2}{*} {$(\ket{b})$} &\multirow{2}{*} {bin $(\ket{b})$}& $(\ket{a})  \approx (\ket{b})$ &$(\ket{a}) > (\ket{b})$ &$(\ket{a})  < (\ket{b})$  & \multirow{2}{*} {Verification}  \\
        
          & &  &   &  & $r_0, r_1$&$r_0, r_1$&$r_0, r_1$&  \\ 
        \hline
        1 &0 & 0 & 0 & 0& 00&-&-& verified \\
        2 &1 & 1 & 0 & 0&-&01&-& verified \\
        3 &0 & 0 & 1 & 1&-&-&10& verified \\
        
        4&3 & 11 & 3 & 11& 00&-&-& verified \\
        5&3 & 11 & 1 & 01&-&01&-& verified \\
        6&1 & 01 & 3 & 11&-&-&10& verified \\
        
        7&7 & 111 & 7 & 111& 00&-&-& verified \\
        8&7 & 111 & 3 & 011&-& 01&-& verified \\
        9&3 & 011 & 7 & 111&-&-&10& verified \\

        10&31 &11111  & 31 & 11111& 00&-&-& verified \\
        11&31 & 11111 & 30 & 11110&-& 01&-& verified \\
        12&30 & 11110 & 31 & 11111&-&-&10& verified \\

        13&120 &1111000  & 120 & 1111000&00&-&-& verified \\
        14&127 &1111111  & 63 & 0111111&-& 01&-& verified \\
        15&100 & 1100100 & 127 & 1111111&-&-&11& verified \\

        16&600 &1001011000  & 600 &1001011000&00&-&-& verified \\
        17&700 &1010111100  & 420 & 0110100100&-& 01&-& verified \\
        18&630 & 1001110110 & 800 & 1100100000&-&-&11& verified \\

        19&1500 &10111011100  & 1500 &10111011100&00&-&-& verified \\
        20&1400 &10101111000  & 200 & 00011001000&-& 01&-& verified \\
        21&560 & 01000110000& 1137 & 10001110001&-&-&11& verified \\
        
        \hline
    \end{tabular}
  \caption{Verification and Demonstration of various Input Sizes and Expected Outcome of GQBSC on Simulators($statevector$ and $qasm$)\label{tab:verification}}
    \end{adjustwidth}
\end{table*}

Validation of the n-bit GQBSC model is established on the metrics of quantum cost, quantum delay, and ancillary bit quantity. These are compared against equivalent QBSC circuits of Wang et. al. \cite{wang2012design}, Al-Rabadi et. al. \cite{al2009closed}, Thapliyal et. al. \cite{thapliyal2010design}, Vudadha et. al. \cite{vudadha2012design}, David et. al. \cite{oliveira2007quantum}, and Xia et. al. \cite{xia2018efficient}. A comparative analysis of these methods is presented in Table.~\ref{tab:comparisonTable}. Here, methods 1-5 are dependent linearly on the input length for the allocation of ancillary qubits, whereas in our case, they remain fixed. This is illustrated in (Fig.~\ref{fig:comparisonall3}a). Method 6 has a lower ancillary bit count but has a higher quantum cost and longer quantum delays.  The quantum cost of our proposed approach is the same as Method 4 if both $\ket{a}$ and $\ket{b}$ are comparable, and slightly larger otherwise by a factor of $(n-1)/2$ (Also illustrated in Fig.~\ref{fig:comparisonall3}b). Here, Method 1 demonstrates exponential cost while the remaining are in linear order. The quantum delay of the proposed approach is illustrated in (Fig.~\ref{fig:GQBSCUtlization}c). Here, method 6  exhibits the highest delays due to the complexity of the comparator design. This is followed by methods 2, 5, and 6 in linear order. The delay of method 1 at the start is close to methods 3 and 4 but as the size increases its delay grows linearly. Our proposed method shows the least delay amongst the linear ordered methods. The least delay is noticeable for methods 3-4 which are of logarithmic order. For small input sizes, our proposed method presents a relatively close delay to that of methods 3-4, but for small input sizes only. 

The GQBSC circuit was tested on IBM System ($ibm\_perth$ and $ibm\_lagos$). The details of the input tests conducted on these systems and their resulting state probabilities are presented in Table ~\ref{tab:QuantumResults}. Given that these systems are equipped with a limited number of 7 qubits, the testing was conducted for a maximum of three-bit numbers.

The table is divided into two sections: one for the results obtained on $ibm\_perth$ and the other for $ibm\_lagos$. It is evident that the results of our proposed GQBSC on $ibm\_perth$ outperform those on $ibm\_lagos$. The proposed circuit was executed for 1024 shots for each input, and the probabilities of the four states are presented in the table.

During the testing, it was observed that as we move toward higher qubit numbers, the output quality begins to deteriorate, and the quantum system generates incorrect outputs.

\begin{table*}
\begin{adjustwidth}{-1in}{-1in} 
\setlength{\tabcolsep}{3pt} 
    \centering
    \begin{tabular}{|c|c|c|c|}
        \hline
        \multirow{2}{*}{Sr.}&Output State& \multirow{2}{*}{Interpretation} & \multirow{2}{*}{Remarks} \\
        &$r_0, r_1$& &\\
            \hline
           1&$00$& {$(\ket{a})  \approx (\ket{b})$} & \texttt{Input-1 == Input-2}\\ 
          2&$01$&$(\ket{a}) > (\ket{b})$  & \texttt{Input-1 > Input-2}\\ 
          3&$10$& $(\ket{a}) < (\ket{b})$  & \texttt{Input-1 < Input-2}\\ 
          4&$11$& $(\ket{a}) < (\ket{b})$  & \texttt{Input-1 < Input-2}\\ 
        \hline
 
    \end{tabular}
  \caption{Interpretation of GQBSC Output States \label{tab:QuantumResultsExplanation}}
    \end{adjustwidth}
\end{table*}

\begin{table*}
\begin{adjustwidth}{-1in}{-1in} 
\setlength{\tabcolsep}{3pt} 
    \centering
    \begin{tabular}{|c|c|c|c|c|c|c|c|c|c|}
        \hline
        \multirow{2}{*}{Sr.}&{$(\ket{a})$} & {$(\ket{b})$} 
        & {$(\ket{a})  \approx (\ket{b})$} &$(\ket{a}) > (\ket{b})$
        
        &$(\ket{a}) < (\ket{b})$ & $(\ket{a}) < (\ket{b})$ &
       { $Highest Prob.$ } & {$Expected$}  & $IBM$ \\
        
          &Input-1& Input-2&$State(0,0)$ &$State(0,1)$&$State(1,0)$& $State(1,1)$ &$State$ &$State$& $Machine$ \\ 
        \hline
            1 & 0 & 0 & 901 & 44 & 74 & 5 & 901 & $True$ & \multirow{16}{*}{$ibm\_perth$}\\
    2 & 0 & 1 & 116 & 10 & 861 & 37 & 861 & $True$ & \\
    3 & 0 & 2 & 110 & 190 & 8 & 716 & 716 & $True$ &  \\
    4 & 0 & 3 & 95 & 218 & 5 & 706 & 706 & $True$ &  \\
    5 & 1 & 0 & 97 & 726 & 20 & 181 & 726 & $True$ &  \\
    6 & 1 & 1 &912&31	&63	&18 & 912 & $True$ &  \\
    7 & 1 & 2 & 127 & 172 & 12 & 713 & 713 & $True$ &  \\
    8 & 1 & 3 & 106 & 241 & 13 & 664 & 664 & $True$ &  \\
    9 & 2 & 0 & 327 & 462 & 7 & 228 & 462 & $True$ &  \\
    10 & 2 & 1 & 176 & 389 & 3 & 456 & 456 & $True$ &  \\
    11 & 2 & 2 & 276 & 505 & 6 & 237 & 505 & \textbf{$False$} &  \\
    12 & 2 & 3 & 155 & 638 & 2 & 229 & 638 & $True$ &  \\
    13 & 3 & 0 & 136 & 310 & 4 & 574 & 574 & $True$ &  \\
    14 & 3 & 1 & 265 & 349 & 3 & 407 & 407 & $True$ &  \\
    15 & 3 & 2 & 175 & 606 & 4 & 239 & 606 & $True$ &  \\
    16 & 3 & 3 & 152 & 657 & 4 & 211 & 657 & \textbf{$False$} &\\
  \hline
        17 & 0 & 0 & 759 & 87 & 166 & 12  & 759 & $True$ & \multirow{16}{*}{$ibm\_lagos$}\\
        18 & 0 & 1 & 160 & 31 & 782 & 51  & 782 & $True$ &  \\
        19 & 0 & 2 & 217 & 154 & 6 & 647  & 647 & $True$ &  \\
        20 & 0 & 3 & 273 & 142 & 9 & 600  & 600 & $True$ &  \\
        21 & 1 & 0 & 197 & 712 & 40 & 75  & 712 & $True$ &  \\
        22 & 1 & 1 & 795 & 83 & 145 & 37  & 795 & $True$ &  \\
        23 & 1 & 2 & 263 & 202 & 8 & 551  & 551 & $True$ &  \\
        24 & 1 & 3 & 228 & 209 & 11 & 567 & 567 & $True$ &  \\
        25 & 2 & 0 & 261 & 446 & 4 & 313 & 446 & $True$ &  \\
        26 & 2 & 1 & 179 & 347 & 31 & 467  & 467 & $True$ &  \\
        27 & 2 & 2 & 335 & 357 & 7 & 325 & 357 & \textbf{$False$} &  \\
        28 & 2 & 3 & 261 & 338 & 9 & 416  & 416 & $True$ &  \\
        29 & 3 & 0 & 236 & 443 & 10 & 338  & 443 & $True$ &  \\
        30 & 3 & 1 & 218 & 308 & 35 & 468  & 468 & $True$ &  \\
        31 & 3 & 2 & 262 & 304 & 14 & 444  & 444 & $True$ &  \\
        32 & 3 & 3 & 203 & 301 & 31 & 489  & 489 &$False$ &  \\
      
        \hline
    \end{tabular}
  \caption{Result of various Input Sizes and Expected Outcome of GQBSC on real Quantum Computer \label{tab:QuantumResults}}
    \end{adjustwidth}
\end{table*}

\begin{table*}
\begin{adjustwidth}{-1in}{-1in} 
\setlength{\tabcolsep}{7pt} 
    \centering
    \begin{tabular}{|c|c|c|c|c|c|}
        \hline
        Method & Reference & Anc. Bits & Quantum Cost & Quantum Delay  \\
        \hline\hline
        1. & Wang et. al. \cite{wang2012design} & $2n$ & $O(n^2)\triangle$ & $O(n^2)$ \\
        2. & Al-Rabadi et. al. \cite{al2009closed} & $6n+1$& $(39n + 9)\triangle$  &$24n+9$ \\
        3. & Thapliyal et. al. \cite{thapliyal2010design} & $4n-3$ & $(18n + 9)\triangle$  & $18\log(2*n)+7$ \\
        4. & Vudadha et. al. \cite{vudadha2012design} & $4n-2$ & $(14n)\triangle$  & $5\log(2*n)+12$ \\
        5. & David et. al. \cite{oliveira2007quantum} & $3n-1$& $(99(n-1)+12)\triangle$  & $20n-1$ \\
        6. & Xia et. al. \cite{xia2018efficient} & 1& $(28n)\triangle$  & $31n+2$ 
       \\\hline
        \multirow{2}{*}{7.} & \multirow{2}{*}{Proposed Comparator} &  \multirow{2}{*}{2} & $14n \triangle;  \quad if a==b$  & $ 4n; \quad  if a==b$ \\
         & &  & $( 14n+(n-1)/2 \triangle  ; \quad if a \neq b$  & $ 4n+(n-1)/2; \quad if a \neq b$ \\
        \hline
    \end{tabular}
    \caption{The ancillary qubit, quantum costs, and quantum delay of proposed method against available methods in literature}
    \label{tab:comparisonTable}
    \end{adjustwidth}
\end{table*}

\pgfplotsset{compat=newest}
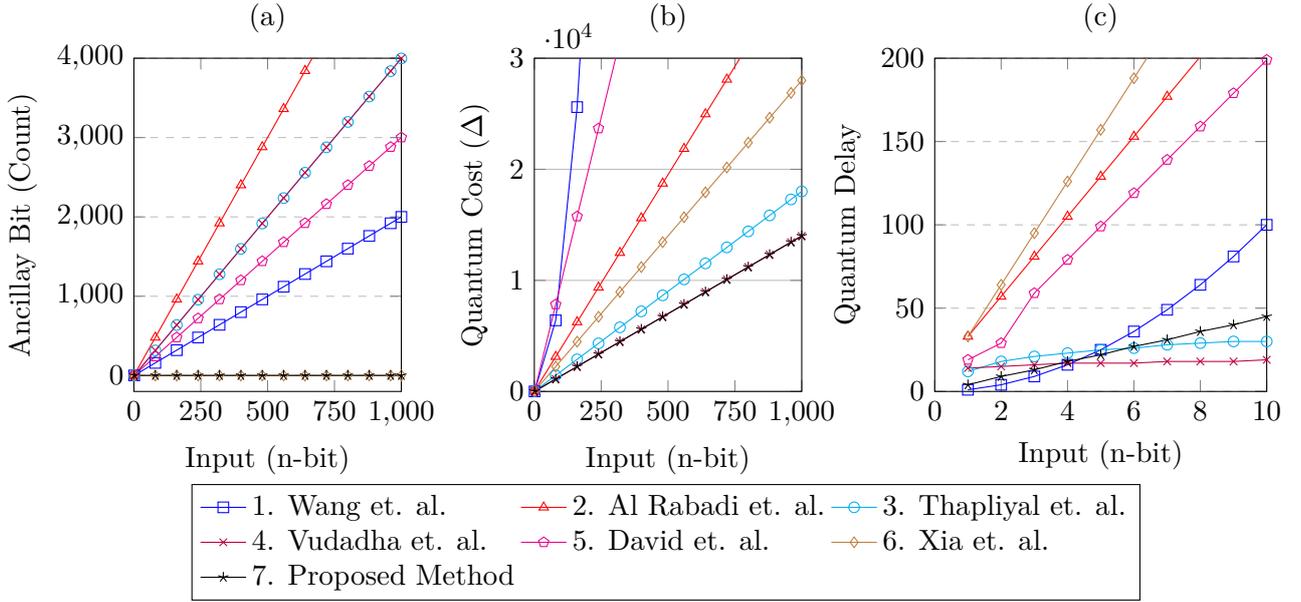
\begin{figure*}
    \begin{tikzpicture}
    \begin{groupplot}[group style={group size= 3 by 1, horizontal sep=1.75cm},height=6cm]
        \nextgroupplot[title=(a),ylabel={Ancillay Bit (Count)},xmin=0, xmax=1000,
        ymin=-200, ymax=4000,
        xtick={0,250,500,750,1000},
        xlabel={Input (n-bit)}, ymajorgrids=true,
        grid style=dashed,
        legend cell align={left},legend to name=zelda,width=.3\textwidth,legend columns=3]
    
    \addplot[color=blue,samples=200,mark=square]
    coordinates {
(1, 2)(80, 160)(160, 320)(240, 480)(320, 640)(400, 800)(480, 960)(560, 1120)(640, 1280)(720, 1440)(800, 1600)(880, 1760)(960, 1920)(1000, 2000)
    };
   \addlegendentry{1. Wang et. al.}

\addplot[
    color=red,samples=100,
    mark=triangle,
    ]
    coordinates {
(1, 7)(80, 481)(160, 961)(240, 1441)(320, 1921)(400, 2401)(480, 2881)(560, 3361)(640, 3841)(720, 4321)(800, 4801)(880, 5281)(960, 5761)(1000, 6001)
    };
   \addlegendentry{2. Al Rabadi et. al.}

\addplot[
    color=cyan,
    mark=o,samples=100
    ]
    coordinates {
(1, 1)(80, 317)(160, 637)(240, 957)(320, 1277)(400, 1597)(480, 1917)(560, 2237)(640, 2557)(720, 2877)(800, 3197)(880, 3517)(960, 3837)(1000, 3997)
    };
    \addlegendentry{3. Thapliyal et. al.}

\addplot[
    color=purple,
    mark=x,samples=100
    ]
    coordinates {
(1, 2)(80, 318)(160, 638)(240, 958)(320, 1278)(400, 1598)(480, 1918)(560, 2238)(640, 2558)(720, 2878)(800, 3198)(880, 3518)(960, 3838)(1000, 3998)
    };
\addlegendentry{4. Vudadha et. al.}

\addplot[
color=magenta,
mark=pentagon,
]
coordinates {
(1, 4)(80, 241)(160, 481)(240, 721)(320, 961)(400, 1201)(480, 1441)(560, 1681)(640,1921)(720, 2161)(800, 2401)(880,2641 )(960, 2881)(1000, 3001)
};
\addlegendentry{5. David et. al.}

\addplot[
    color=brown,
    mark=diamond,samples=100
    ]
    coordinates {
(1, 1)(80, 1)(160, 1)(240, 1)(320, 1)(400, 1)(480, 1)(560, 1)(640, 1)(720, 1)(800, 1)(880, 1)(960, 1)(1000, 1)
};
\addlegendentry{6. Xia et. al.}

\addplot[
color=black,
mark=star,samples=100
]
coordinates {
(1, 2)(80, 2)(160, 2)(240, 2)(320, 2)(400, 2)(480, 2)(560, 2)(640, 2)(720, 2)(800, 2)(880, 2)(960, 2)(1000, 2)

};
\addlegendentry{7. Proposed Method}

        \nextgroupplot[title=(b),xlabel={Input (n-bit)},
ylabel={Quantum Cost ($\Delta$)}, xmin=0, xmax=1000, xtick={0,250, 500, 750, 1000}, ymin=0, ymax=30000 , ymajorgrids=true, width=.3\textwidth
]

\addplot[
 color=blue,
    mark=square,
    ]
    coordinates {
(1, 1)(80, 6400)(160, 25600)(240, 57600)(320, 102400)(400, 160000)(480, 230400)(560, 313600)(640, 409600)(720, 518400)(800, 640000)(880, 774400)(960, 921600)(1000, 1000000)
    };

\addplot[
    color=red,
    mark=triangle,
    ]
    coordinates {
(1, 48)(80, 3129)(160, 6249)(240, 9369)(320, 12489)(400, 15609)(480, 18729)(560, 21849)(640, 24969)(720, 28089) (800, 31209) (880, 34329) (960, 37449)(1000, 39009)
    };

\addplot[
    color=cyan,
    mark=o,
    ]
    coordinates {
(1, 27)(80, 1449)(160, 2889)(240, 4329) (320, 5769)(400, 7209)(480, 8649)(560, 10089)(640, 11529) (720, 12969) (800, 14409) (880, 15849)(960, 17289) (1000, 18009)
    };

\addplot[
    color=purple,
    mark=x,
    ]
    coordinates {
(1, 14)(80, 1120)(160, 2240)(240, 3360)(320, 4480)(400, 5600)(480, 6720)(560, 7840)(640, 8960)(720, 10080)(800, 11200)(880, 12320)(960, 13440) (1000, 14000)
    };

\addplot[
color=magenta,
mark=pentagon,
]
coordinates {
(1, 12)(80, 7833)(160, 15753)(240, 23673)(320, 31593)(400, 39513)(480, 47433)(560, 55353)(640,63273)(720, 71193)(800, 79113)(880,87033 )(960, 94953)(1000, 98913)
};

\addplot[
    color=brown,
    mark=diamond,
    ]
    coordinates {
 (1, 28)(80, 2240)(160, 4480)(240, 6720)(320, 8960)(400, 11200)(480, 13440)(560, 15680) (640, 17920)  (720, 20160)(800, 22400)(880, 24640)  (960, 26880) (1000, 28000)
};

\addplot[
color=black,
mark=star,
]
coordinates {
(1, 14) (80, 1120)  (160, 2240)  (240, 3360) (320, 4480) (400, 5600)(480, 6720) (560, 7840)  (640, 8960) (720, 10080) (800, 11200)  (880, 12320) (960, 13440)  (1000, 14000)
};

        \nextgroupplot[title=(c), ylabel={Quantum Delay}, xlabel={Input (n-bit)},
    xmin=0, xmax=10,
    ymin=0, ymax=200,
    legend pos= north west,
    ymajorgrids=true,
    grid style=dashed, width=.35\textwidth]

   \addplot[
 color=blue,
    mark=square,
    ]
    coordinates { (1, 1) (2, 4) (3, 9) (4, 16) (5, 25) (6, 36) (7, 49) (8, 64) (9, 81) (10, 100)
    };

\addplot[
    color=red,
    mark=triangle,
    ]
    coordinates {
(1, 33) (2, 57) (3, 81) (4, 105) (5, 129) (6, 153) (7, 177) (8, 201) (9, 225) (10, 249)
    };

\addplot[
    color=cyan,
    mark=o,
    ]
    coordinates {
(1, 12) (2, 18) (3, 21) (4, 23) (5, 25) (6, 26) (7, 28) (8, 29) (9, 30) (10, 30)
    };

\addplot[
    color=purple,
    mark=x,
    ]
    coordinates {
(1, 14) (2, 15) (3, 16) (4, 17) (5, 17) (6, 17) (7, 18) (8, 18) (9, 18) (10, 19)
    };

\addplot[
color=magenta,
mark=pentagon,
]
coordinates {
(1,19) (2, 29) (3, 59) (4, 79) (5, 99) (6, 119) (7, 139) (8, 159) (9, 179) (10, 199)
};

\addplot[
    color=brown,
    mark=diamond,
    ]
    coordinates {
(1, 33) (2, 64) (3, 95) (4, 126) (5, 157) (6, 188) (7, 219) (8, 250) (9, 281) (10, 312)
};

\addplot[
color=black,
mark=star,
]
coordinates {
(1, 4) (2, 9) (3, 13) (4, 18) (5, 22) (6, 27) (7, 31) (8, 36) (9, 40) (10, 45)
};

    \end{groupplot}
    \node[] at(7,-2) {\pgfplotslegendfromname{zelda}};
\end{tikzpicture}
    \caption{Graphical plot of the number of (a) ancillary bits, (b) quantum cost, and (c) quantum delay of the proposed method against comparable methods of literature.\label{fig:comparisonall3}}

\end{figure*}
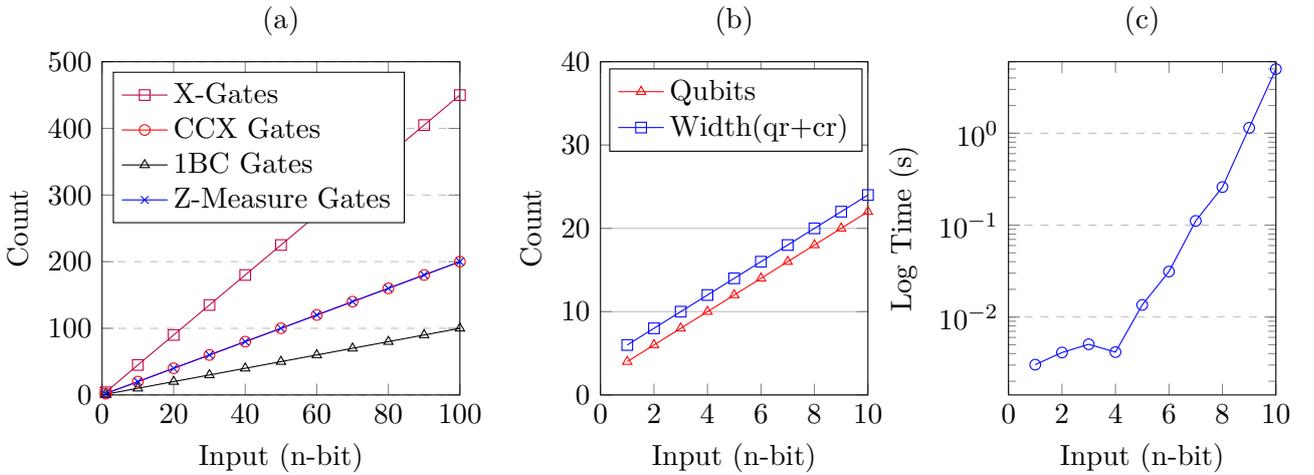
\begin{figure*}
    \begin{tikzpicture}
    \begin{groupplot}[group style={group size= 3 by 1, horizontal sep=1.85cm},height=6cm]
        \nextgroupplot[title=(a),xlabel={Input (n-bit)},ylabel={Count},
		xmin=0, xmax=100,
		ymin=0, ymax=500,
        xtick={0,20,40,60,80,100},
        ytick={0,100,200,300,400,500},
        ymajorgrids=true,
        grid style=dashed,
        legend pos= north west,
        width=.37\textwidth,legend columns=1,legend cell align={left}]
   		\addplot[color=purple,mark=square] coordinates { (1, 4) (10, 45)  (20, 90)  (30, 135)  (40, 180)(50, 225) (60, 270) (70, 315) (80, 360)  (90, 405) (100, 450)};
		\addlegendentry{X-Gates}
		
		\addplot[color=red,mark=o] coordinates {	(1, 2) (10, 20)  (20, 40)  (30, 60)  (40, 80)  (50, 100) (60, 120)  (70, 140)  (80, 160) (90, 180)  (100, 200)};
		\addlegendentry{CCX Gates}
		
		\addplot[color=black,mark=triangle] coordinates { (1, 1) (10, 10)  (20, 20)  (30, 30)(40, 40)(50, 50)(60, 60) (70, 70) (80, 80) (90, 90) (100, 100)
		};
		\addlegendentry{1BC Gates}
		
		\addplot[color=blue,mark=x] coordinates {(1, 2) (10, 20) (20, 40) (30, 60)  (40, 80)  (50, 100)  (60, 120) (70, 140)  (80, 160)  (90, 180) (100, 200)
		};
		\addlegendentry{Z-Measure Gates}

        \nextgroupplot[title=(b),xlabel={Input (n-bit)},
ylabel={Count}, xmin=0, xmax=10, ymin=0, ymax=40 , ymajorgrids=true, width=.3\textwidth,legend cell align={left}
]
		\addplot[
		color=red,
		mark=triangle,
		]
			coordinates{
		(1, 4) (2, 6) (3, 8) (4, 10) (5, 12) (6, 14) (7, 16) (8, 18) (9, 20) (10, 22)
	};
		\addlegendentry{Qubits}
		\addplot[
		color=blue,
		mark=square,
		]
		coordinates {(1, 6) (2, 8) (3, 10) (4, 12) (5, 14) (6, 16) (7, 18) (8, 20) (9, 22) (10, 24)
		};
		\addlegendentry{Width(qr+cr)}
        \nextgroupplot[title=(c),
    xlabel={Input (n-bit)},
		ylabel={Log Time (s)},
		xmin=0, xmax=10,
		ymin=-1, ymax=6,
    legend pos= north west,
    ymajorgrids=true,
    grid style=dashed, width=.3\textwidth, legend cell align={left},ymode=log]
  		\addplot[
	color=blue,
	mark=o,
		]
		coordinates {(1, 0.00303) (2, 0.00411) (3, 0.00505) (4, 0.00415) (5, 0.01354) (6, 0.03126) (7, 0.11116) (8, 0.25942) (9, 1.14506) (10, 4.98984)
		};

    \end{groupplot}

\end{tikzpicture}
    \caption{Graphical plot of (a) number of X, CCX, 1BC, and measurement gates, (b) number of Qubits and the width of quantum and classical registers, and (c) run time in seconds, for the proposed method \label{fig:GQBSCUtlization}}
 \end{figure*}

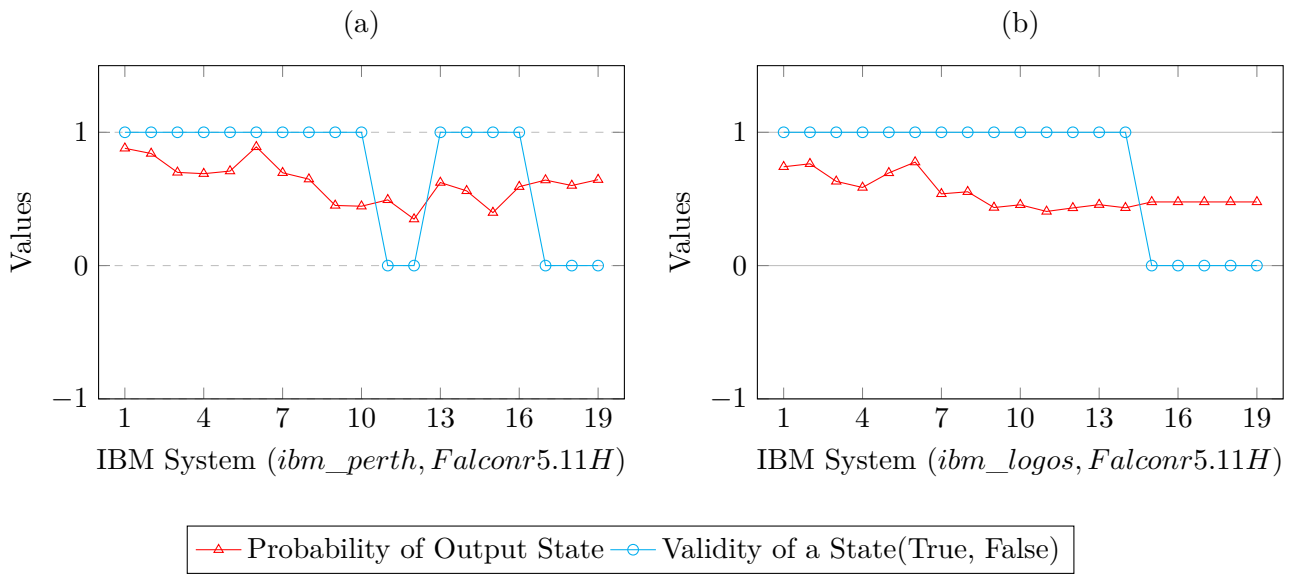
\begin{figure*}
    \begin{tikzpicture}
    \begin{groupplot}[group style={group size= 3 by 1, horizontal sep=1.75cm},height=6cm]
        \nextgroupplot[title=(a),ylabel={Values},xmin=0, xmax=20,
        ymin=-1, ymax=1.5,
        xtick={1,4,7,10,13,16,19},
        xlabel={IBM System $(ibm\_perth,Falcon r5.11H)$}, ymajorgrids=true,
        grid style=dashed,
        legend cell align={left},legend to name=zelda,width=.5\textwidth,legend columns=3]
    

\addplot[
    color=red,samples=100,
    mark=triangle,
    ]
    coordinates {
(1, 0.879882813) (2, 0.840820313) (3, 0.69921875) (4, 0.689453125) (5, 0.708984375) (6, 0.890625) (7, 0.696289063) (8, 0.6484375) (9, 0.451171875)(10, 0.4453125) (11, 0.493164063) (12, 0.348632813) (13, 0.623046875) (14, 0.560546875) (15, 0.397460938) (16, 0.591796875) (17, 0.641601563) (18, 0.600585938) (19, 0.64453125)
    };
   \addlegendentry{Probability of Output State}

\addplot[
    color=cyan,
    mark=o,samples=100
    ]
    coordinates {
(1, 1) (2, 1) (3, 1) (4, 1) (5, 1) (6, 1) (7, 1) (8, 1) (9, 1) (10, 1) (11, 0) (12, 0) (13, 1) (14, 1) (15, 1) (16, 1) (17, 0) (18, 0) (19, 0)
    };
    \addlegendentry{Validity of a State(True, False)}






 \nextgroupplot[title=(b),ylabel={Values},xmin=0, xmax=20,
        ymin=-1, ymax=1.5,
        xtick={1,4,7,10,13,16,19},
        xlabel={IBM System $(ibm\_logos,Falcon r5.11H)$}, ymajorgrids=true,
        width=.5\textwidth]


\addplot[
    color=red,samples=100,
    mark=triangle,
    ]
    coordinates {
(1, 0.741210938)(2, 0.763671875) (3, 0.631835938) (4, 0.5859375) (5, 0.6953125) (6, 0.776367188) (7, 0.538085938) (8, 0.553710938) (9, 0.435546875) (10, 0.456054688) (11, 0.40625) (12, 0.432617188) (13, 0.45703125) (14, 0.43359375) (15, 0.477539063) (16, 0.477539063) (17, 0.477539063) (18, 0.477539063) (19, 0.477539063)
    };

\addplot[
    color=cyan,
    mark=o,samples=100
    ]
    coordinates {
(1, 1) (2, 1) (3, 1) (4, 1) (5, 1) (6, 1) (7, 1) (8, 1) (9, 1) (10, 1) (11, 1) (12, 1) (13, 1) (14, 1) (15, 0) (16, 0) (17, 0) (18, 0) (19, 0)
    };



    \end{groupplot}
    \node[] at(7,-2) {\pgfplotslegendfromname{zelda}};
\end{tikzpicture}
    \caption{Graphical plot of the number of (a) and (b) show the different result of proposed method on IBM Systems .\label{fig:QuantumReults}}

\end{figure*}

The comparator design is further evaluated through the number of gate operations and circuit complexity and tested against bitstring inputs of various sizes. This is presented in Fig.~\ref{fig:GQBSCUtlization}. Here, (a) presents the design with respect to the number of X, CCX, 1BC, and Measurement gates, (b) presents the number of qubits and accumulative width of the quantum and classical registers. Fig.~\ref{fig:GQBSCUtlization}c represents the run-time duration when executed on a simulator. Notably, it demonstrates a relatively constant runtime at the outset, but as $n$ grows larger, it exhibits an increase.

The resource analysis reveals that the proposed quantum comparator $n$ qubit for holding the information and $2$ ancillary qubit for storing the result of comparison, where $n$ represents the size of the binary numbers being compared. This linear scaling suggests that the comparator can handle larger numbers with a manageable increase in resource requirements. The potential advantages of quantum comparators over classical counterparts include parallel processing, reduced computational complexity, and possible speedup in specific applications.

We now present a behavior analysis of our proposed n-bit quantum bit string operator by delineating all the gate operations that it encompasses in the algorithm. The representation allows us to establish a composition-based mathematical representation of the input bitstrings and the quantum output states. This is expressed as in (Equ. ~\ref{equation:GeneralizeRepresentation}):
\begin{widetext}
\begin{equation}
\begin{array}{l}
\ket{r_0} = 
\begin{cases}
    \begin{array}{l}
    \ket{r_0}_i \oplus \left(\ket{a_1}_i \land \ket{b_0}_i\right),
    \end{array} & \text{if } \text{n}= 1 \\
    \\
    \begin{array}{l}
    \left[ \ket{r_0}_i \oplus \left(\ket{a_1}_i \land \ket{b_0}_i \right)\right] \\
    \qquad\qquad \oplus \left[\ket{a_1}_{i+1} \land \ket{b_0}_{i+1}\right],
    \end{array}
     & \text{if } \text{n}>1, \ket{r_0}_i  = \ket{0},  \ket{r_1}_i = \ket{0}, \ket{r_1}_{i+1} \neq \ket{1}\\
    \\
    \begin{array}{l}
    \left[\ket{r_0}_i \oplus \left(\ket{a_1}_i\land \ket{b_0}_i \right)\right]\\
    \qquad\qquad \oplus [\ket{a_1}_{i+1} \land \ket{b_0}_{i+1}] \cdot [\text{X} (\ket{r_0}_{i+1})],
    \end{array} & \text{if } \text{n}  > 1\ket{r_0}_i  = \ket{0}, \ket{r_1}_i  =  \ket{0},  \ket{r_1}_{i+1} = \ket{1}
\end{cases}
\\
\\
\ket{r_1} = \begin{cases}  
    \begin{array}{l}
    \ket{r_1}_i \oplus (\ket{a_0}_i  \land \ket{b_1}_i) 
    \end{array}, & \qquad\qquad\quad\text{if } \text{n} = 1 \\
    \\
    \begin{array}{l}
    \left[\ket{r_1}_i \oplus (\ket{a_0}_i \land \ket{b_1}_i \right] \\
    \qquad\qquad \oplus \left[\ket{a_1}_{i+1} \land \ket{b_0}_{i+1}\right],
    \end{array} 
    & \qquad\qquad\quad \text{if } \text{n} > 1, \ket{r_0}_i = \ket{0}, \ket{r_1}_i = \ket{0}, \ket{r_1}_{i+1} \neq \ket{1}
\end{cases}
\end{array}
\label{equation:GeneralizeRepresentation}
\end{equation}
\end{widetext}
where $n$ is the bit string length, $\ket{r_0}$ and $\ket{r_1}$ contain the final results of the comparator, $i < n$ is the current iteration, and consequently $i+1$ is the next iteration, and any $\ket{\psi}_i$ represents the intermediate results of a quantum state with respect to the iterator $i$. The final output $\ket{r_0}$ and $\ket{r_1}$ is conditional and depends on the bit string length, and the intermediate result states approximation to either $\ket{0}$ or $\ket{1}$. By applying these conditions, we can achieve the same outcome as that obtained from our proposed method discussed in Algo.~\ref{algo:GeneralizeAlgorithm} or illustrated in Figs.~\ref{fig:1bit}-\ref{fig:5bit}.

In Fig. ~\ref{fig:QuantumReults}, we have included some statistics of the proposed GQBSC  method applied to the IBM System ($ibm\_perth$ and $ibm\_lagos$), both of which have a system size of 7 qubits. These graphs clearly illustrate that as the input size approaches the maximum capacity of the qubit system, the performance of the proposed state begins to deteriorate, and the probability of the output state eventually declines for comparable input sizes.

\section{Conclusion}

Quantum bit string comparators are essential to many quantum algorithms as it provides for means to compare two quantum states. In this study, we have investigated the resource requirements and scalability of our proposed quantum comparator. The analysis demonstrates that our approach efficiently compares input states of varying sizes, with an upper bound resource scaling of n+2 qubits. Holistically, our proposed approach fares better when compared to other methods on the basis of ancillary qubits, quantum cost, and quantum delay. These findings contribute to the ongoing development of quantum algorithms. The proposed approach is extensively tested, verified, and validated against other methods.

%
%
%
%
%

\end{document}